\documentstyle[psfig]{mn}

\def\apj{ApJ}
\def\apjl{ApJ}

\def\mnras{MNRAS}
\def\aap{A\&A}

\def\pasj{PASJ}

\def\pasp{PASP}

\def\ref{\par\noindent\hang}
\def\spose#1{\hbox to 0pt{#1\hss}}
\def\approxlt{\mathrel{\spose{\lower 3pt\hbox{$\sim$}}
        \raise 2.0pt\hbox{$$<$$}}}
\def\approxgt{\mathrel{\spose{\lower 3pt\hbox{$\sim$}}
        \raise 2.0pt\hbox{$>$}}}

\def\multleft#1{\hbox to size{\vbox {\halign {\lft{##}\cr #1}}\hfill}\par}
\def\multright#1{\hbox to size{\vbox {\halign {\rt{##}\cr #1}}\hfill}\par}

\def\today{\ifcase\month\or January\or February\or March\or April\or May\or
      June\or July\or August\or September\or October\or November\or December\fi
      \space\number\day, \number\year}
\def\s{\hbox{\phantom{5}}}      

\def\boxit#1{\vbox{\hrule\hbox{\vrule\kern3pt\vbox{\kern3pt
          #1 \kern3pt}\kern3pt\vrule}\hrule}}

\def\cm{{\rm\thinspace cm\thinspace}}

\def\erg{{\rm\thinspace erg\thinspace}}
\def\eV{{\rm\thinspace eV\thinspace}}

\def\K{{\rm\thinspace K\thinspace}}
\def\keV{{\rm\thinspace keV\thinspace}}
\def\km{{\rm\thinspace km\thinspace}}

\def\Mpc{{\rm\thinspace Mpc\thinspace}}

\def\s{{\rm\thinspace s\thinspace}}

\def\pcmcu{\hbox{$\cm^{-3}\,$}}

\def\ergcmps{\hbox{$\erg\cm\ps\,$}}

\def\ergps{\hbox{$\erg\s^{-1}\,$}}

\def\kmps{\hbox{$\km\s^{-1}\,$}}

\def\ps{\hbox{$\s^{-1}\,$}}

\def\kmpspMpc{\hbox{$\kmps\Mpc^{-1}$}}

\begin{document}
\hsize=6truein

\title{An extended multi-zone model for the MCG$-$6-30-15 warm absorber}

\author[]
{\parbox[]{6.in} {R. Morales$^1$, A.C.~Fabian$^1$ and  C.S. Reynolds$^2$ \\
\footnotesize
1. Institute of Astronomy, Madingley Road, Cambridge CB3 0HA \\
2. JILA, University of Colorado, Campus Box 440, Boulder, CO
80309-0440 USA\\ }}                                            
\maketitle

\begin{abstract} 
  The variable warm absorber seen with {\em ASCA} in the X-ray
  spectrum of MCG$-$6-30-15 shows complex time behaviour in which the
  optical depth of OVIII anticorrelates with the flux whereas that of
  OVII is unchanging. The explanation in terms of a two zone absorber
  has since been challenged by {\em BeppoSAX} observations. These
  present a more complicated behaviour for the OVIII edge. We
  demonstrate here that the presence of a third, intermediate, zone
  can explain all the observations. In practice, warm absorbers are
  likely to be extended, multi-zone regions of which only part causes
  directly observable absorption edges at any given time.
\end{abstract}

\begin{keywords} galaxies: active $-$ galaxies: individual:
  MCG$-$6-30-15 $-$ galaxies: Seyfert $-$ X-rays: galaxies.
\end{keywords}

\section{INTRODUCTION} 

X-ray absorption by partially ionized, optically thin material, along
the line of sight of the central engine, the so called {\em warm
  absorber}, is one of the prominent features in the X-ray spectrum of
many AGN \cite{reyno97}. The presence of such gas was first postulated
in order to explain the unusual form of the X-ray spectrum of QSO MR
2251-178 \cite{hal84}.  A direct confirmation of the existence of
circumnuclear ionized matter came from {\em ASCA}, which for the first
time was able to resolve the OVII and OVIII absorption edges (at rest
energies of 0.74 and 0.87 \keV, respectively) in the X-ray spectra of
the Seyfert 1 galaxy MCG$-$6-30-15 \cite{fab94}. Systematic studies of
warm absorbers in Seyfert 1 galaxies with {\em ASCA} have shown their
ubiquity, being detected in half of the sources \cite{reyno97}.

Warm absorbers were usually described by single zone, photoionization
equilibrium models. Under this assumption, simple variability patterns
were expected: increasing ionization of the matter when the source
brightens. This was not found by Otani et al. \shortcite{ota96} during
their long-look {\em ASCA} observation of the nearby (z=0.008) Seyfert
1 galaxy MCG$-$6-30-15. The source varied with large amplitude on
short time scales ($10^4$ s or so). The depth of the OVII edge, on the
contrary, stayed almost constant, while that of the OVIII edge was
anticorrelated with the flux. To explain the behaviour of the OVII and
OVIII edges, the authors adopted a multizone model in which the OVII
and OVIII edges originate from spatially distinct regions. OVII ions
in the region responsible for the OVII edge, the outer absorber, have
a long recombination timescale (i.e. weeks or more), whereas the OVIII
edge arises from more highly ionized material, the inner absorber, in
which most oxygen is fully stripped.  The recombination timescale for
the OIX ions in the inner absorber is of the order of $10^4 s$ or
less. A decrease in the primary ionizing flux is then accompanied by
the recombination of OIX to OVIII, giving the observed variation in
the OVIII edge depth.

Orr et al. \shortcite{orr97} raised the possibility of a more complex
warm absorber.  During their MCG$-$6-30-15 {\em BeppoSAX} observation
the depth of the OVII edge was marginally consistent with being
constant, whereas the optical depth for OVIII, $\tau(OVIII)$,
exhibited significant variability. The authors claimed that its large
value during epoch 1\footnote{The epochs in the {\em BeppoSAX}
  observation are chronologically numerated (i.e. number 1 corresponds
  to the first epoch of the observation, etc.).} ($1.7\pm0.5$,
$1\sigma$ uncertainty) was inconsistent with the values at all other
epochs.  They also pointed out that the value of $\tau(OVIII)$ during
epoch 5 (a low luminosity state) did not show the expected
anticorrelaction with the ionizing flux .

In both Otani et al. \shortcite{ota96} and Orr et
al.\shortcite{orr97}, $\tau(OVIII)$ was plotted versus count rate.  In
order to compare {\em ASCA} and {\em BeppoSAX} observations, these two
observational results for $\tau(OVIII)$ have been plotted versus
luminosity\footnote{The conversion from count rate to luminosity has
  been obtained using $H_o=50$ \kmpspMpc, $q_0=0$ and PIMMS
  (http://heasarc.gsfc.nasa.gov/Tools/w3pimms.html).} in 0.1-10 \keV
in figure 1 (instead of versus count rate). The conversion from count
rate to luminosity is different for each apparatus and therefore, the
comparison can not be made using count rate, but luminosity.
\begin{figure}
\centerline{\psfig{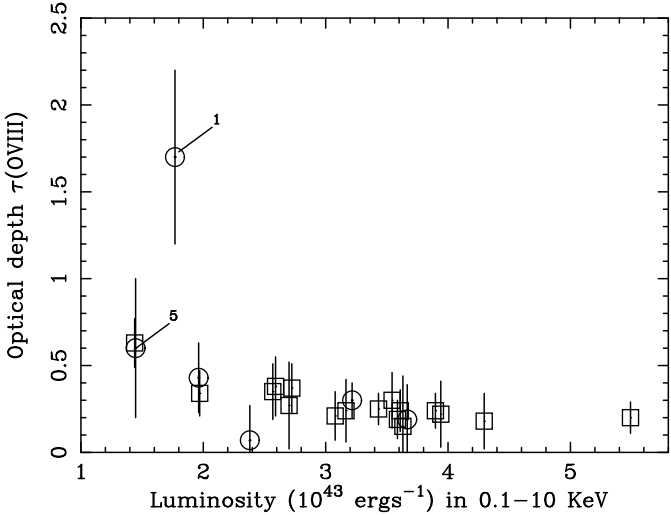}}
\caption{Comparison {\em ASCA} (square) - {\em BeppoSAX} (circle)
  observational results for the optical depth of OVIII, $\tau(OVIII)$,
  versus luminosity in 0.1-10 \keV. Note the only discrepant point in
  the {\em BeppoSAX} observation labelled as 1 and the very good
  agreement for point 5.}
\end{figure}
The only discrepancy between both sets of data is the result for epoch
1 (labelled 1 in figure 1)\footnote{This epoch was at the very
  beginning of the {\em BeppoSAX} observation. In a private
  communication, A. Orr notes that this high value for the
  optical depth for OVIII could not be due to instrumental effects,
  since no similar behaviour, neither at the beginning nor during the
  observation has been observed in any other {\em BeppoSAX} target
  yet.}. Note also that there is no disagreement at all with any other
point (Orr et al. mentioned epoch 5 as problematic, but in this plot
it appears consistent with the rest of the
Otani et al. data).

In this paper we present a simple photoionization model that accounts
for the experimental results of both {\em ASCA} and {\em BeppoSAX}
observations. Section 2 describes the code used to model the inner
warm absorber. The application to the MCG$-$6-30-15 warm absorber is
presented in Section 3. The extension of the multi-zone model for the
warm absorber (i. e. the third component) is addressed in Section
4. Finally our conclusions are discussed in Section 5.

\section{TIME DEPENDENT PHOTOIONIZATION CODE}

In order to reproduce both {\em ASCA} and {\em BeppoSAX} observations,
the state of the inner absorber has been modelled using a time
dependent photoionization code for oxygen \cite{rey96}. This code treats the
material in the inner absorber as containing only ions of oxygen in an
otherwise completely ionized hydrogen plasma.  A given oxygen ion is
assumed to be in one of the 9 states corresponding to its ionization
level. Different excitation levels of a given ionization state are not
treated. The total abundance of oxygen is fixed at $7.41\times
10^{-4}$\cite{gre89} relative to hydrogen. Both ions and electrons are
assumed to be in local thermal equilibrium (LTE) with a common
temperature T and the plasma is assumed to be strictly optically-thin
at all frequencies (i.e. radiative transfer processes are not
considered).

A point source of ionizing radiation is placed at a distance $R$ from
this material with a frequency dependent luminosity $L_{\nu}$ (total
luminosity $L$). In the calculations presented here, the ionizing
continuum is taken to be a power-law with a photon index $\Gamma=2$
extending from $\nu_{min}$ to $\nu_{max}$: i.e. we take

\begin{equation}
L_{\nu}=\frac{L}{\Lambda\nu}
\end{equation}
where $\Lambda=\ln(\nu_{max}/
\nu_{min})$. The upper and lower cutoff frequencies are chosen such
that $h\nu_{max}=100$ \keV, and $\Lambda=10$, corresponding to
$h\nu_{min}\approx4.5$ \eV.

The physical processes included in the code are: photoionization,
Auger ionization, collisional ionization, radiative recombination,
dielectronic recombination, bremsstrahlung, Compton scattering and
collisionally-excited O$\lambda$1035 resonance line cooling. Given
these processes, the ionization and thermal structure of the plasma is
evolved from an initial state assumed to have a temperature of
$T=10^5$ \K, and equal ionization fractions in each state.

The ionization structure of the oxygen is governed by 9 equations of
the form:
\begin{equation}
\frac{dn_i}{dt}=\sum_{prod.}\delta_{prod.}-\sum_{dest.}\delta_{dest.}
\end{equation}
where $n_i$ is the number density of state-i, $\delta$ stands for the
ionization/recombination rate per unit volume, the first summation on
the right hand side (RHS) is over all ionization and recombination
mechanisms that produce state-i and the second summation is over all
such mechanisms that destroy state-i.

The thermal evolution of the plasma is determined by the local
heating/cooling rate and the macroscopic constraint. The net heating
rate per unit volume is given by:
\begin{equation}
\frac{dQ}{dt}=\sum_{heat}\Delta_{heat}-\sum_{cool}\Delta_{cool}
\end{equation}
where $\Delta$ stands for the heating/cooling rate per unit volume,
the first summation on the RHS is over all heating processes and the
second term is over all cooling processes. For a {\em constant
  density} plasma, we have $du=dQ$ where $u$ is the internal energy
per unit volume and is given by:
\begin{equation} 
u=\frac{3}{2}n_ek_BT.
\end{equation}
Thus, the rate of change of temperature is related to the
heating/cooling rate via:
\begin{equation}
\frac{dT}{dt}=\frac{2}{3n_ek_B}\frac{dQ}{dt}.
\end{equation}

The state of the plasma is evolved using the ionization equations
represented by (2) and the energy equation (3). A simple step-by-step
integration of these differential equation is performed. The time step
for the integration process is allowed to dynamically change and is
set to be $0.1\times t_{sh}$ where $t_{sh}$ is the shortest relevant
timescale\footnote{At each step the ionization, collisional and
  recombination timescales of each oxygen ionization state is
  calculated. The shortest of these timescales is taken to be
  $t_{sh}$.}. In the case of a constant ionizing luminosity, it is
found that the system always evolves to an equilibrium state, as
expected.

The model has been compared against the photoionization code {\sc
  cloudy}\cite{ferla96} (version 9004, \cite{ferl98}) in the case of a
plasma in photoionization equilibrium.  Figures 2 and 3 report this
comparison.There is a qualitative agreement between the model (solid
line) and {\sc cloudy} (dashed line) in figure 2 for the case of a
pure hydrogen/oxygen plasma.  The main goal of this paper is to
reproduce the time variability of the oxygen edges, and, this figure
shows how our model mirrors quite well the behavior of this element.
{\sc cloudy} includes many more physical processes and realistic
elemental abundances and, as seen in
figure 3, the agreement is not so good when more elements are
considered\footnote{Realistic elemental abundances can be important
  both for the ionization structure and for the heating/cooling rate
  of the plasma, and since features of the
  stability curve result from peaks in the heating and cooling
  functions (which are in turn associated with particular ionic
  species), the comparison presented in figure 2 is expected to present
  a much better agreement. The consideration of realistic abundances
  is beyond the scope of this paper and is currently being studied.}.
\begin{figure}
\centerline{\psfig{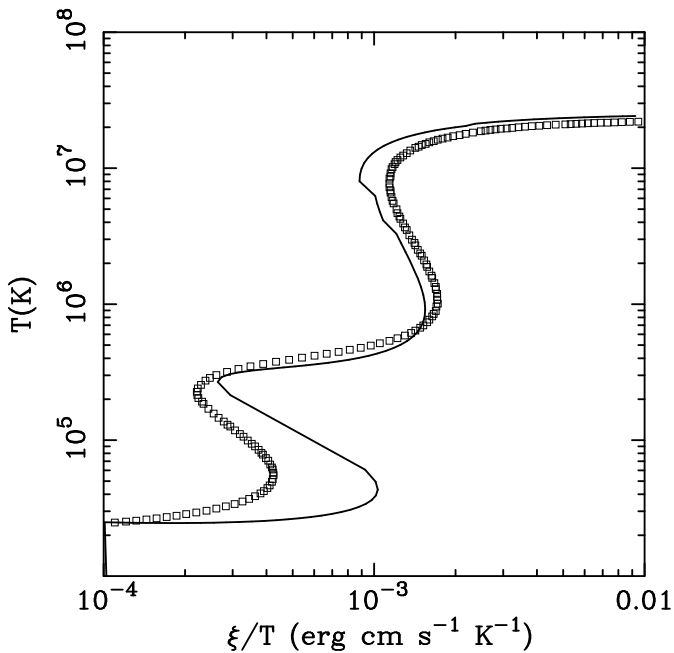}}
\caption{Comparison of the model (solid curve) with {\sc cloudy}
(dashed line) of a pure hydrogen/oxygen plasma.}          
\end{figure}
\begin{figure}
\centerline{\psfig{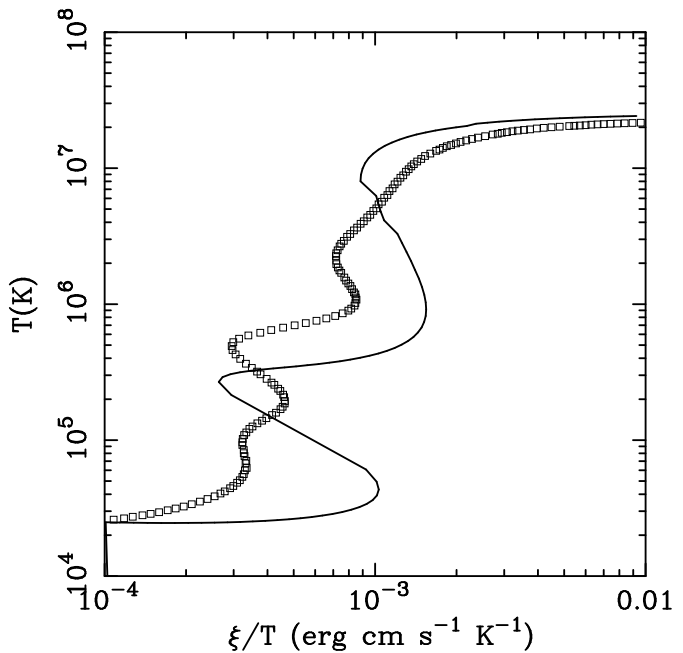}}
\caption{Comparison of the model (solid curve) with {\sc cloudy}
(dashed line).}
\end{figure}
We have also investigated the reaction of a constant density plasma to
an inverted top hat function for the light-curve (i.e. initially
$T=10^5$ \K, equal ionization fractions in each state, the luminosity
is $L=3\times10^{43}$ \ergps and held constant. The system is allowed
to achieve an equilibrium at an ionization parameter
$\xi=\frac{L}{nR^2}=75$ \ergcmps, where $n$ is the density of the warm
plasma and $R$ is the distance of the slab of the plasma from the
ionizing source of radiation with isotropic ionizing luminosity $L$.
The luminosity is then halved for a short period, $\Delta t$, and then
back to its initial value) . During these changes, the state of the
plasma is recorded as a function of time. This choice of $\xi$ places
the inner warm absorber a region where $f_{O8}\propto L^{-1}$ (i.e. in
this regime $f_{O8}$ is expected to be inversely proportional to $\xi$
or equivalently to $L$). The investigation has been carried out for
different values of $n$ and $R$ keeping $n>2\times10^7$ \pcmcu and
$R<1.4\times10^{17}$ \cm (i.e. the constraints given by Otani et
al.\shortcite{ota96} for the inner warm absorber).  Fig.  4 shows an
example of the results obtained. As expected, $f_{O8}$ increases when
$L$ is halved, showing a more noticeable increment when $L$ is halved
over a longer period.
\begin{figure}
\centerline{\psfig{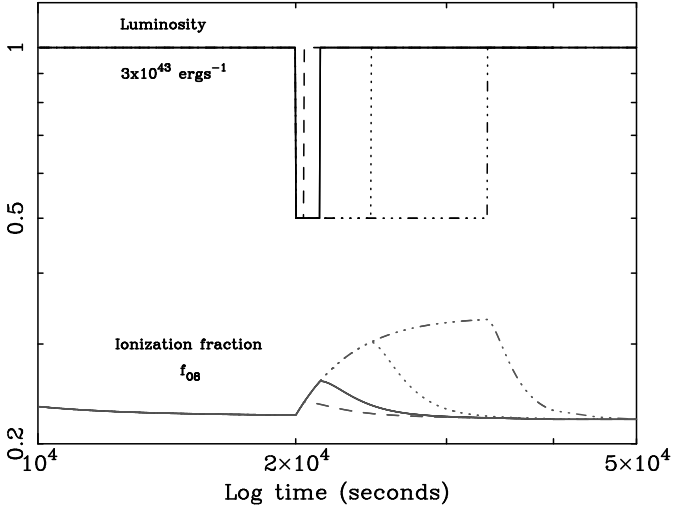}}
\caption{Time behaviour of the ionization fraction for OVIII,
  $f_{O8}$, for different light-curves. The plasma has ionization
  parameter $\xi=75$ \ergcmps for luminosity $L=3\times10^{43}$
  \ergps, electron density $n_e=3\times10^7$ \pcmcu and it is situated
  at $1.15\times10^{17}$ \cm from the ionizing source. The luminosity
  $L$ is halved over an interval $\Delta t$ equal to: 450(dashed
  line), 1350(solid line), 4500(dot line) and 13500(dot-dot-dot-dashed
  line) seconds. The corresponding ionization fraction, $f_{O8}$, is
  plotted below.}
\end{figure}

\section{APPLICATION TO THE MCG$-$6-30-15 WARM ABSORBER}

To reproduce the observational data using our model, it is required to
obtain the value of the optical depth for OVIII once the ionization
fraction $f_{O8}$ has been calculated by the code.  Considering the
cross section for OVIII, $\sigma_8$, as constant along the
line-of-sight and using the fact that the material contains only ions
of oxygen in an otherwise completely ionized plasma, the optical depth
for OVIII, $\tau(OVIII)$, can be written as:
\begin{equation}
\tau(OVIII)=\frac{\sigma_8f_{O8}\Delta R n_e}{\frac{1}{Abun}+\sum
\limits_{i=1}^9 f_{Oi}\times (i-1)}
\end{equation}
where $\sigma_8=10^{-19} cm^2$
\cite{ost89}, $\Delta R$ is the line-of-sight
distance through the ionized plasma, $n_e$ is the electron density,
$Abun$ is the oxygen abundance relative to hydrogen\cite{gre89} and
$f_{Oi}$ is the ionization fraction for i ionization species for oxygen.

The light-curve used to reproduced the {\em ASCA} observation is given
by Otani et al. \shortcite{ota96}. For the {\em BeppoSAX} observation,
the light-curve has been defined as a step function with a constant
value for the luminosity $L$ over the different time periods given by
Orr et al. \shortcite{orr97}.  The constant value for $L$ over each
period is chosen to be that given in Orr et al. (1997, Fig.
3)\nocite{orr97} plus a period of $2\times 10^4$ \s previous to epoch
1 with $L=1.8\times10^{43}$ \ergps (a standard value for the
luminosity during the observation). Different parameters for the warm
absorber have been investigated using the constraints given by Otani
et al.  \shortcite{ota96}.  With the only exception of point 1 (see
figure 1), a general good agreement for all other experimental points
is found.

Following a suggestion by Orr et al. \shortcite{orr97}, we have
modified the light-curve for the {\em BeppoSAX} observation including
a previous epoch to it for which the luminosity is much lower (the
range of values used is $10^{40}-10^{43}$ \ergps)\footnote{An example
  of a Seyfert 1 galaxy exhibiting an unusual low flux state (a
  decrease of more than one order of magnitude in luminosity) is
  presented in Uttley et al. \shortcite{utt99}.}. The reason for using
this very low value for the luminosity is that in a regime for which
$f_{O8}\propto L^{-1}$, the highest values $f_{O8}$ are expected for
low values of $L$. However, the suggested explanation does not
reproduce the high value of $\tau (OVIII)$ for epoch 1 in the {\em
  BeppoSAX} observation.  Even for those values of $\xi$ that give a
maximum for $f_{O8}$ (i.e.  $\xi\approx 50$ \ergcmps), the ionization
fraction for O(VIII) is still too low to account for the high $\tau
(OVIII)$ value at epoch 1.

\section{A THIRD WARM ABSORBER COMPONENT}

The model we propose to explain both {\em ASCA} and {\em BeppoSAX}
observations incorporates a new zone for the warm absorber. Let warm
absorber 1 $\equiv$ WA1 be the inner warm absorber which parameters
are: $R<1.4\times10^{17}$ \cm, $n>2\times10^7$ \pcmcu and $\Delta
R\simeq 10^{14}$ \cm. The outer warm absorber will be warm absorber 2
$\equiv$ WA2 with parameters $R>3\times10^{18}$ \cm, $n<2\times10^5$
\pcmcu and $\Delta R\simeq 10^{14}$ \cm. In our model a warm absorber
3 $\equiv$ WA3 is situated between WA1 and WA2. The WA3 radius and
density will have values between those of WA1 and WA2. Therefore,
while WA1 and WA2 respond on timescales of hours and weeks
respectively, WA3 is expected to respond to variations in the ionizing
flux on timescales of days. WA3 would have an ionization parameter
$\xi$ of the order of 500 \ergcmps for $L=3\times10^{43}$ \ergps.
Hence the ionization fraction for this value of $L$ is too low to be
detected. Only when the luminosity of the source is sufficiently low
(i.e. $\xi\approx 50$ \ergcmps), WA3 reveals its presence by
contributing to the total optical depth for OVIII. The medium between
WA1 and WA2 would be constituted by a continuum of warm absorbers:
clouds with different densities situated at different radii. Only some
of them happen to be at the radius and have densities that efficiently
absorbs the OVIII edge energy (i.e. most of them are undetectable). In
Baldwin et al. \shortcite{bal95} a model for the BLR is presented, in
which individual BLR clouds can be thought of as machines for
reprocessing radiation.  As long as there are enough clouds at the
correct radius and with the correct gas density to efficiently form a
given line, the line will be formed with a relative strength which
turns out to be very similar to the one observed. Similarly in our
model, only WA1 and WA2 are detectable for the ordinary values of the
ionizing flux. Only for the case of a state of low luminosity, WA3
will be unmasked. Other zones, as yet unseen, may be present.

Assuming then the presence of WA3 and also an epoch of low luminosity
previous to the {\em BeppoSAX} observation, the expected WA3 behaviour
would be:

i) when $L\approx 0.4\times10^{43}$ \ergps (previous to epoch 1), the
ionization parameter $\xi\approx 50$ \ergcmps, giving a high value for
$f_{O8}$. This period lasts for approximately $10^5$ \s, so the plasma
has time to recombine.

ii) when $L\approx (1,5)\times 10^{43}$ \ergps (i.e. during the
observation), then $\xi\approx 150,750$ \ergcmps. For these high
values of $\xi$, oxygen is practically fully stripped and therefore
there is a very small contribution to the optical depth for OVIII.

The range of parameters investigated for WA3 is $R=(2,8)\times
10^{17}$ cm, $n=5\times10^5,10^7$ \pcmcu and $\Delta R$ in the
interval that gives a column density for WA3 approximately equal to
$3\times 10^{22}$ \pcmcu. After taking into account the soft X-ray
absorption due to WA1, the response of WA3 has been calculated and an
example of the results obtained is presented in figures 5 and 6, where
the general good agreement is also extended to point 1. 

WA3 has also been modelled using {\sc cloudy} and we found a drop in
the transmitted portion of the incident continuum at $\approx 7.8$
\keV, of approximately $1\%$ (i.e. undetectable for current
instruments). The coronal lines strength has also been checked using
{\sc cloudy}\footnote{See Sect. 3.3 \cite{Ferg97} for a discussion of
  the state of the coronal lines atomic data used in {\sc cloudy}.}.
The ratio of the modelled to observed fluxes for WA3 are all below
$0.1\times f_c$, where $f_c$ is the covering fraction\footnote{Porquet
  et al. \shortcite{Porq99} give some restrictions on the density of
  the WA in order to avoid producing coronal line equivalent widths
  larger than observed. Although WA3 presents no problems at all for
  any $f_c$, WA2 does, unless a low value of $f_c$ is considered. This
  possibility is currently being investigated.}.

Finally, the contribution to $\tau$(OVII) from each component for our
model has been calculated and, as expected, we found that WA2 is the
main responsible for the OVII edge (its contribution makes 96\% of the
total value of $\tau$(OVII)). 
\begin{figure}
\centerline{\psfig{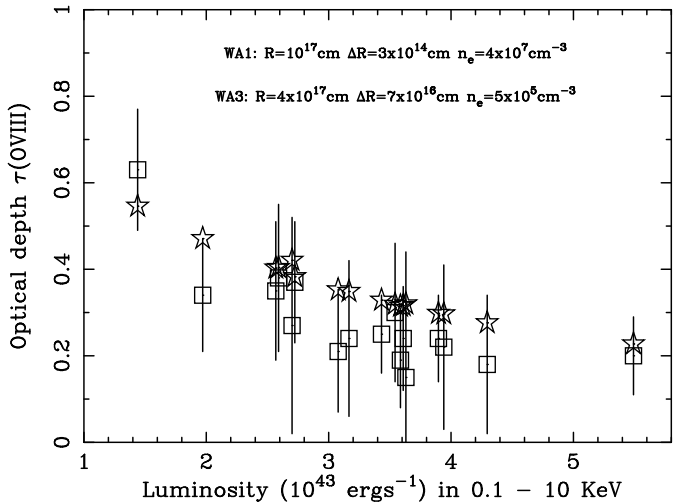}}
\caption{Comparison {\em ASCA} data (square) with the model
  computations (star) for the following warm absorber parameters: WA1:
  distance to the ionizing source, $R=1.0\times10^{17}$ \cm,
  line-of-sight distance through the warm absorber, $\Delta
  R=3.0\times 10^{14}$ \cm, and electron density $n_e=4.0\times10^7$ \pcmcu.
  WA3: $R=4.0\times10^{17}$ \cm, $\Delta R=7.0\times 10^{16}$ \cm, and
  $n_e=5.0\times 10^5$ \pcmcu.}
\end{figure}
\begin{figure}
\centerline{\psfig{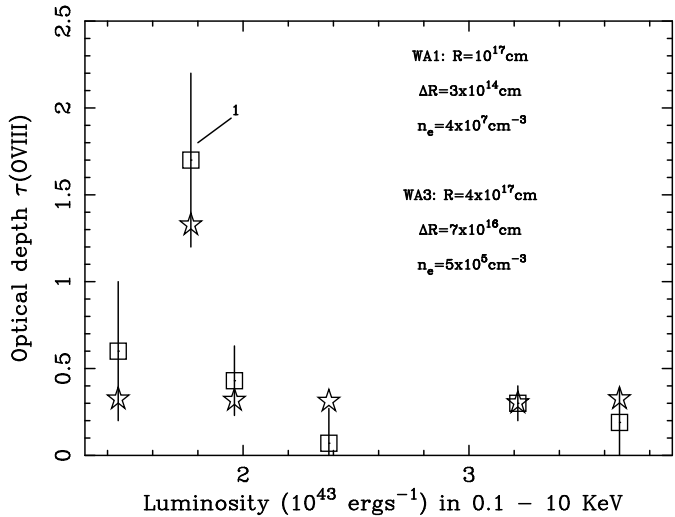}}
\caption{Comparison {\em BeppoSAX} data (square) with the model
  computations (star) for the following warm absorber parameters: WA1:
  distance to the ionizing source, $R=1.0\times10^{17}$ \cm,
  line-of-sight distance through the warm absorber, $\Delta
  R=3.0\times 10^{14}$ \cm, and electron density $n_e=4.0\times10^7$ \pcmcu.
  WA3: $R=4.0\times10^{17}$ \cm, $\Delta R=7.0\times 10^{16}$ \cm, and
  $n_e=5.0\times 10^5$ \pcmcu.}
\end{figure}

\section{DISCUSSION}
Warm absorbers have been the subject of extensive studies during the
last decade. Such regions are not spatially resolved, and all the
available information about their geometry is obtained from analysis
of the variability of the oxygen edges. The explanation we offer for
the time variability of the MCG$-$6-30-15 warm absorber during both
{\em ASCA} and {\em BeppoSAX} observations does not invoke complex
processes, but a very simple photoionization model together with the
presence of a multi-zone warm absorber. This would be constituted by a
continuum of clouds at different radii and different densities, such
that only some of them contribute to the total optical depth for OVIII
depending on the value of the luminosity.

As a final remark note how in our model WA3 is much more volume
filling than WA1 ($\Delta R/R\geq 0.1$ for WA3 while $\Delta R/R\leq 10^{-3}$
for WA1). This suggests that WA3 could be part of the inter-cloud
medium of WA1.
\section{ACKNOWLEDGMENTS}
This work has been supported by PPARC and Trinity College (R.M.) and by
the Royal Society (A.C.F.). C.S.R. thanks support from Hubble
Fellowship grant HF-01113.01-98A.  This grant was awarded by the Space
Telescope Institute, which is operated by the Association of
Universities for Research in Astronomy, Inc., for NASA under contract
NAS 5-26555.  C.S.R. also thanks support from NASA under LTSA grant
NAG5-6337.

\bibliographystyle{mnras}

\end{document}